# HIGH-SPIN STATES IN AND AROUND DOUBLY-MAGIC NUCLEI*

R. Broda

H. Niewodniczański Institute of Nuclear Physics, Kraków, Poland

The study of high-spin states in regions of doubly-magic nuclei performed with the use of deep-inelastic heavy ion reactions is reviewed. New and tentative results concerning high-spin states in the $^{48}$Ca and yrast structures in $^{47}$Ca, $^{47}$K, $^{49}$Ca and $^{49}$Sc isotopes are presented. The status of the high-spin state study in the region of $^{132}$Sn and $^{208}$Pb is outlined, including discussion of recently obtained results in the $^{208}$Pb core and the $^{206}$Hg two-proton-hole nucleus.

## 1. Introduction

The unusual richness of the scientific activity of Professor Zdzisaw Szymaski, as well as his extraordinary charisma, made strong impact on the nuclear physics community in Kraków. When I considered, which would be the most proper way to pay respect to the memory of our beloved friend, admired teacher and great compatriot, I came to conclusion, that being experimentalist I should try to select some exciting new experimental results for the presentation at this conference. In any experimental research process there is an initial stage when new results are rather tentative and particularly their interpretation is very incomplete. Significant part of my talk will be devoted to such cases, however I shall also present few results which are already well elaborated and published. This has two implications:

The unpublished new results come from the data analysis of recent weeks and there was no time for any consultation with other participants of research. Therefore I have to take my own responsibility in presenting them publicly; the list of many co-authors who contributed to these results is given in the acknowledgment.

* Invited talk presented at the "High Spin Physics 2001" NATO Advanced Research Workshop, dedicated to the memory of Zdzisław Szymański, Warsaw, Poland, February 6–10, 2001

Presenting new results I shall indicate difficulties related to their interpretation and I shall emphasize our hope to get dedicated support from theorists. Such demonstration might be the best way to show how much we appreciate such support and how much we owe to Zdzisaw for his heritage, which is so impressively represented by the activity of Warsaw nuclear physics theory groups.

The high-spin state study since many years is an important part of nuclear physics research. It is focussed mainly on collective phenomena and interplay between collective and shell model aspects of nuclear structure, notably the study of rotation of deformed nuclei is one of the main objectives. In many cases also spherical nuclei are studied, where high spins are acquired by the coupling of angular momentum of individual nucleons and the yrast line is largely determined by the energetically favored multi-particle configurations. These studies have great impact on testing and improving the quantitative shell model description. The extraordinary purity often observed for such high spin states allows to extract important ingredients for shell model calculations. It is obvious that the doubly-magic nuclei and their closest neighbors are the most appropriate objects for such attempts. Yet, until recently, the experiments devoted to the high spin state study of doubly-magic nuclei regions faced serious difficulties due to the lack of suitable reactions populating these states in a way which would enable successful spectroscopic analysis.

The situation has changed when large germanium multi-detector arrays were constructed, thereby providing efficiency and resolving power satisfactory to unfold very complex gamma coincidence spectra. Since several years we are using these excellent tools in high statistics gamma ray coincidence experiments performed with thick targets and various systems of heavy ions colliding at energies exceeding by 10 to 20% the Coulomb barrier. Very complex deep-inelastic reactions taking place at these energies are usually accompanied by massive exchange of nucleons and often produce nuclei that cannot be reached in standard fusion evaporation processes. Moreover the final product nuclei are populated up to rather high spin states with yields which enable yrast spectroscopy analysis in nuclei previously inaccessible for such study. The technique is based solely on the coincidence data analysis of discrete gamma transitions emitted from the product nuclei stopped in the target material; gamma transitions emitted in flight remain unobserved due to the Doppler broadening. Within the last decade numerous spectroscopic results were obtained on yrast structures of nuclei located at and beyond the neutron-rich edge of the beta stability valley [1-5]. Very early attempts were also made to search for high spin states in regions of doubly magic nuclei [6-8]. In this talk I shall focus on new results obtained in regions of $^{48}$Ca, $^{132}$Sn and $^{208}$Pb doubly-magic nuclei.

## 2. High-spin states in the $^{48}$Ca core nucleus and lowest yrast excitations in its 1 particle (hole) neighbors

Scarce information on yrast states in the doubly-magic $^{48}$Ca nucleus and its closest neighbors comes predominantly from simple light particle scattering and transfer reaction experiments [9,10]. Consequently only very few high spin states were identified, often with tentative spin-parity assignments and approx. 10 keV accuracy for the excitation energy.

Our new results come essentially from the experiment performed with the GASP array at the INFN Legnaro Laboratory, in which the 1.2 mg/cm2 $^{48}$Ca target backed by a thick $^{208}$Pb material was bombarded with the $^{48}$Ca beam of 210 MeV energy from the ALPI Linac accelerator. Earlier experiment performed by us with the EUROBALL array for the same symmetric $^{48}$Ca+$^{48}$Ca system, but at much lower beam energy of 140 MeV , did not show any significant yields for the expected deep-inelastic reaction products. Surprisingly, in spite of the 30 MeV excess of the beam energy over the Coulomb barrier, the gamma coincidence data contained only events arising from the fusion evaporation products, with a very small contribution which could be attributed to quasi-elastic peripheral processes. On the other hand the data from the additional $^{48}$Ca+$^{208}$Pb experiment performed with the GAMMASPHERE array at the Argonne NL provided useful check of some identifications. In all, the data analysis of the $^{48}$Ca+210MeV $^{48}$Ca run provided cleanest and most significant results which are presented below.

In the analysis an important starting point was the pre-selection of low multiplicity (up to fold 7 ) events, which contained nearly all gamma coincidences of $^{48}$Ca-like products and rejected fully those arising from the Zr, Y, Sr fusion evaporation residues, as well as reduced significantly events originating from nuclear reactions with unavoidable oxygen contamination in the target. The use of the "low-multiplicity filter" may sound paradoxical, as our aim was to search for high spin states in the $^{48}$Ca and neighboring nuclei, nevertheless such choice turned out to be very profitable to achieve our goal.

The gamma coincidence analysis of high spin states in the $^{48}$Ca core nucleus was particularly simple since the level scheme was well known in its lowest part. Fig. 1 shows the scheme of yrast levels in the $^{48}$Ca including new states identified from our data. All states located below 6 MeV excitation energy were known previously. The energetically favored simple particle-hole excitations are likely to dominate the structure of these lowest lying states. The sequence of positive parity states with spin values I=2,3,4,5 must involve largely the excitation of the $f_{7/2}$ neutron across the N=28 gap to the $p_{3/2}$ shell. On the other hand the promotion of the proton across the Z=20 gap, from one of the filled $s_{1/2}$ or $d_{3/2}$orbital to the empty $f_{7/2}$ shell, gives rise to negative parity states of which the I=3,4,5 are near yrast states. Here, only the observation of the 758 keV gamma transition connecting the $4^-$ and $4^+$ states, and the suggestion of the $5^+$ spin parity assignment for the 5146 keV level, based on its strong yrast population and gamma decay, come from the present work. All other states shown in the Fig. 1 level scheme are new; unfortunately they remain at present without any spin-parity assignments. Nevertheless, it is clear that most of them are near yrast states with spin values higher than I=5 previously observed in the doubly-magic 48Ca nucleus. It is also clear that all of them must involve two-particle two-hole core excitations, which is manifested by their large energy separation from lower lying states.

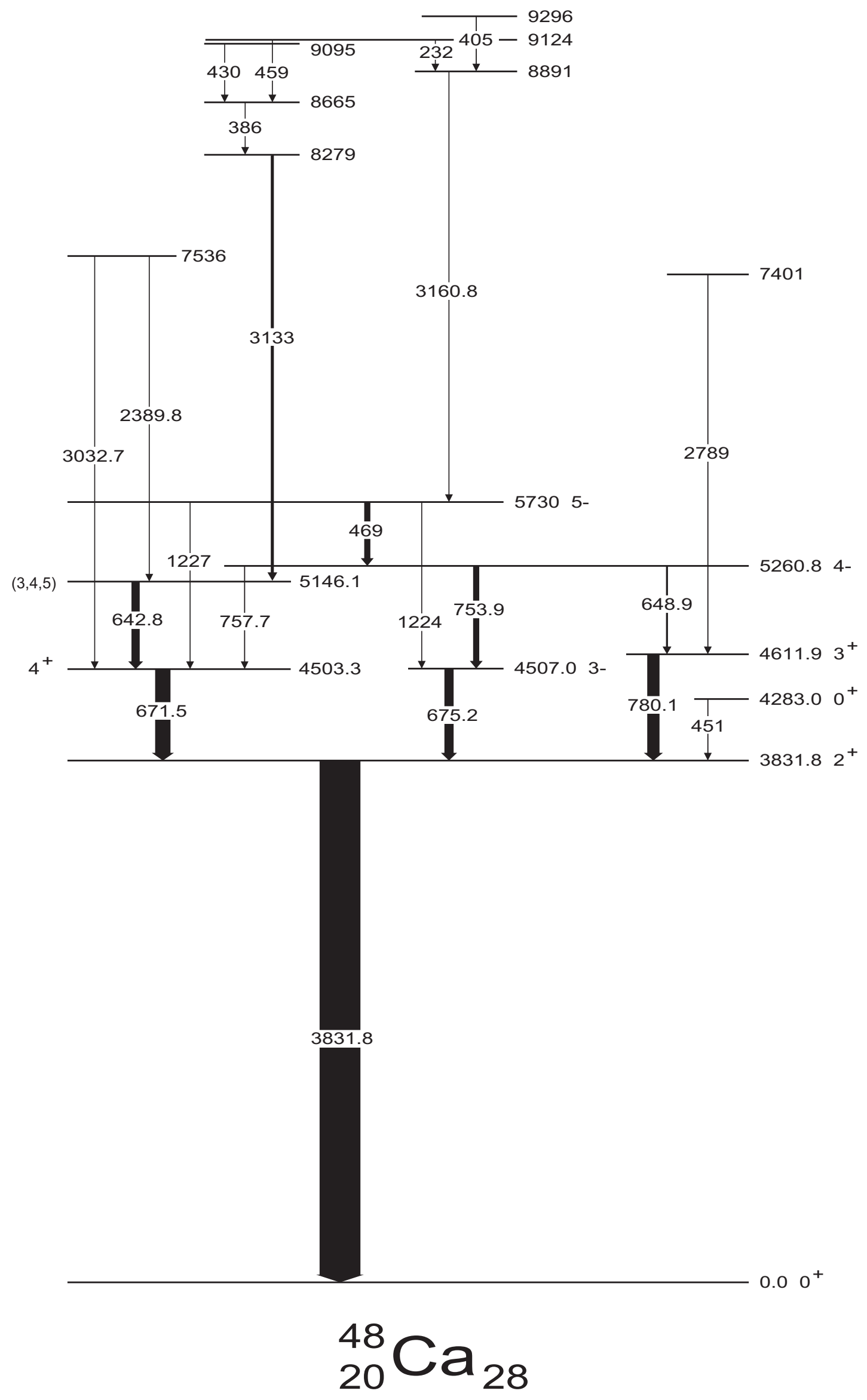

Fig. 1. Level scheme of the $^{48}$Ca obtained in the $^{48}$Ca+ 210 MeV $^{48}$Ca experiment.

The coincidence spectrum obtained with the gate set on the most intense $2^+ \rightarrow 0^+$ 3832 keV transition, shown in Fig. 2, demonstrates the quality of the data. Apart from gamma lines representing transitions from higher lying states of $^{48}$Ca, as well as from another $^{48}$Ca reaction partner, other observed transitions can be attributed to the $^{47}$Ca and $^{46}$Ca nuclei produced in more violent collisions which are followed by correspondingly one- or two-neutron evaporation. Whereas the $^{46}$Ca transitions are well known, those arising from the $^{47}$Ca are now identified or confirmed basing on previous knowledge of this nucleus. In the high energy part of Fig. 2 spectrum the 2014, 3999, 3933 and 3562 keV lines clearly belong to the $^{47}$Ca and represent transitions which establish earlier observed levels [10]. The observed gamma coincidences allowed to construct a simple level scheme of $^{47}$Ca as shown in Fig. 3. The previously known and unambiguously assigned levels at 2014 keV (3/2-), 2578 keV (3/2+) and 2599 keV (1/2+) are fully confirmed and naturally interpreted as arising from correspondingly promotion of an unpaired neutron to the $p_{3/2}$orbital above N=28, and filling of the $f_{7/2}$ hole with $d_{3/2}$ and $s_{1/2}$ sub-shell neutrons below N=20. However, tentative spin-parity assignments suggested for other states arising from the coupling of the $f_{7/2}$ neutron hole with $^{48}$Ca core excitations have to be questioned. Whereas the $13/2^+$ assignment for the 3999 keV level seems to be likely, its observed gamma decay excludes $5/2^-$,$7/2^-$assignment for the 3933 keV state. The observed gamma branching would suggest E1 (65 keV), M2 (437 keV), E3 (3999 keV) transitions, which then assigns the 3933 keV state as $11/2^-$, and confirms the earlier suggested $9/2^-$ spin-parity for the 3562 keV level. Although the energies of higher lying states at 4403 and 4811 keV fit well those of the earlier observed levels, their suggested $1/2^-$, $3/2^-$ assignments are completely excluded; the tentative15/2 and 17/2 assignments are much more likely.

Fig. 3 includes also our new level scheme of the one-neutron $^{49}$Ca nucleus. The identification of the 3357 and 660 keV most intense transitions was based on observed cross-coincidences with $^{47}$Ca, $^{46}$Ca and $^{45}$Ca transitions and absence of any transitions from the $^{48}$Ca. The three lowest levels at 3357, 3867 and 4017 keV are identified with earlier observed states at energies systematically lower by approx 6 keV [10]. Although there is little doubt that this are the same states, the earlier suggested spin-parity assignments have again to be questioned. In particular the $3/2^-$ assignment for the 3866 keV state is practically excluded. At this moment all assignments for earlier known states and for the new higher lying levels in $^{49}$Ca would be highly speculative.

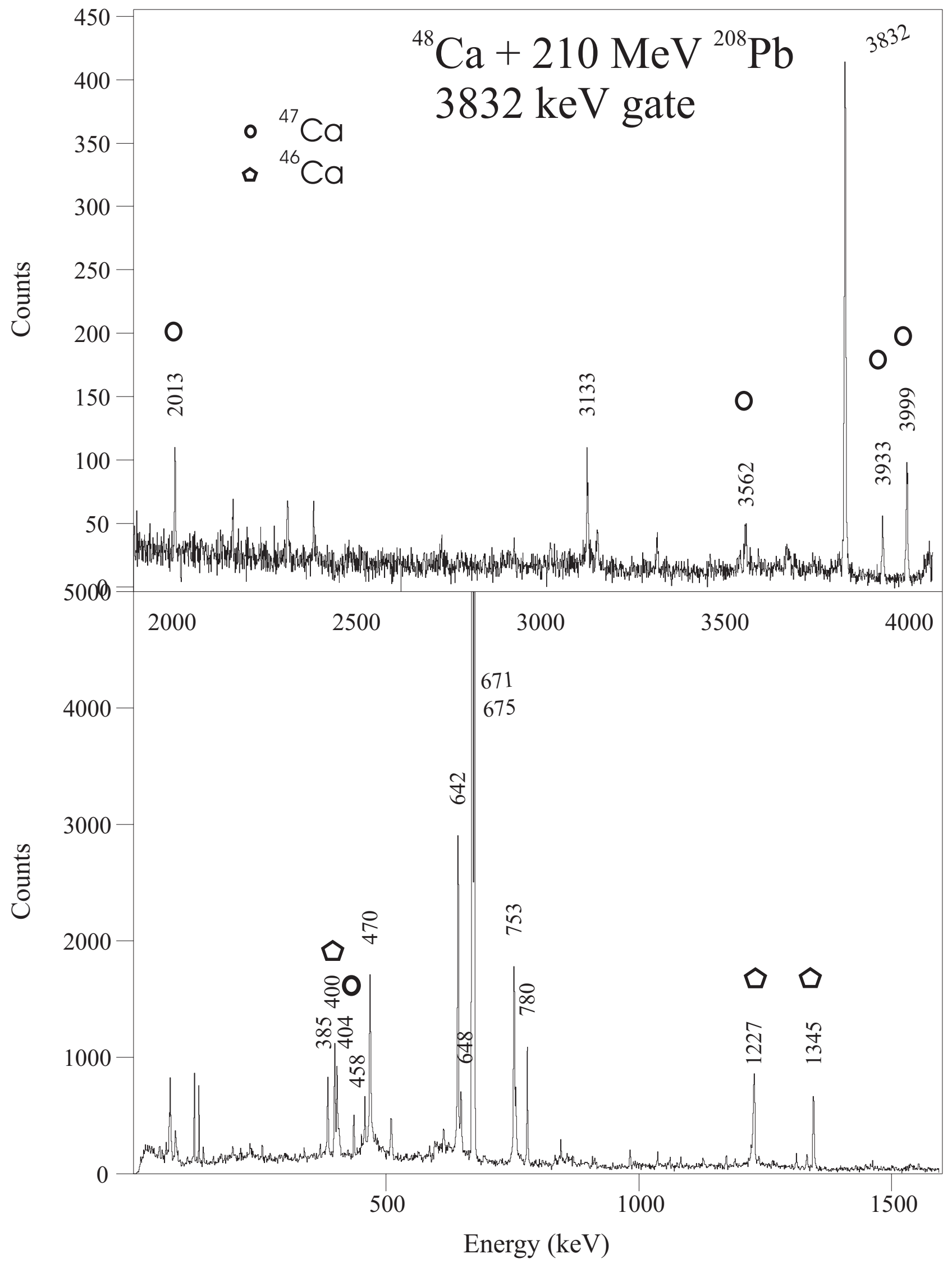

Fig. 2. Gamma coincidence spectrum with the 3832 keV $2^+ \to 0^+$ transition in $^{48}$Ca. Transitions with indicated energies are from $^{48}$Ca, those marked by symbols are identified with $^{47}$Ca and $^{46}$Ca.

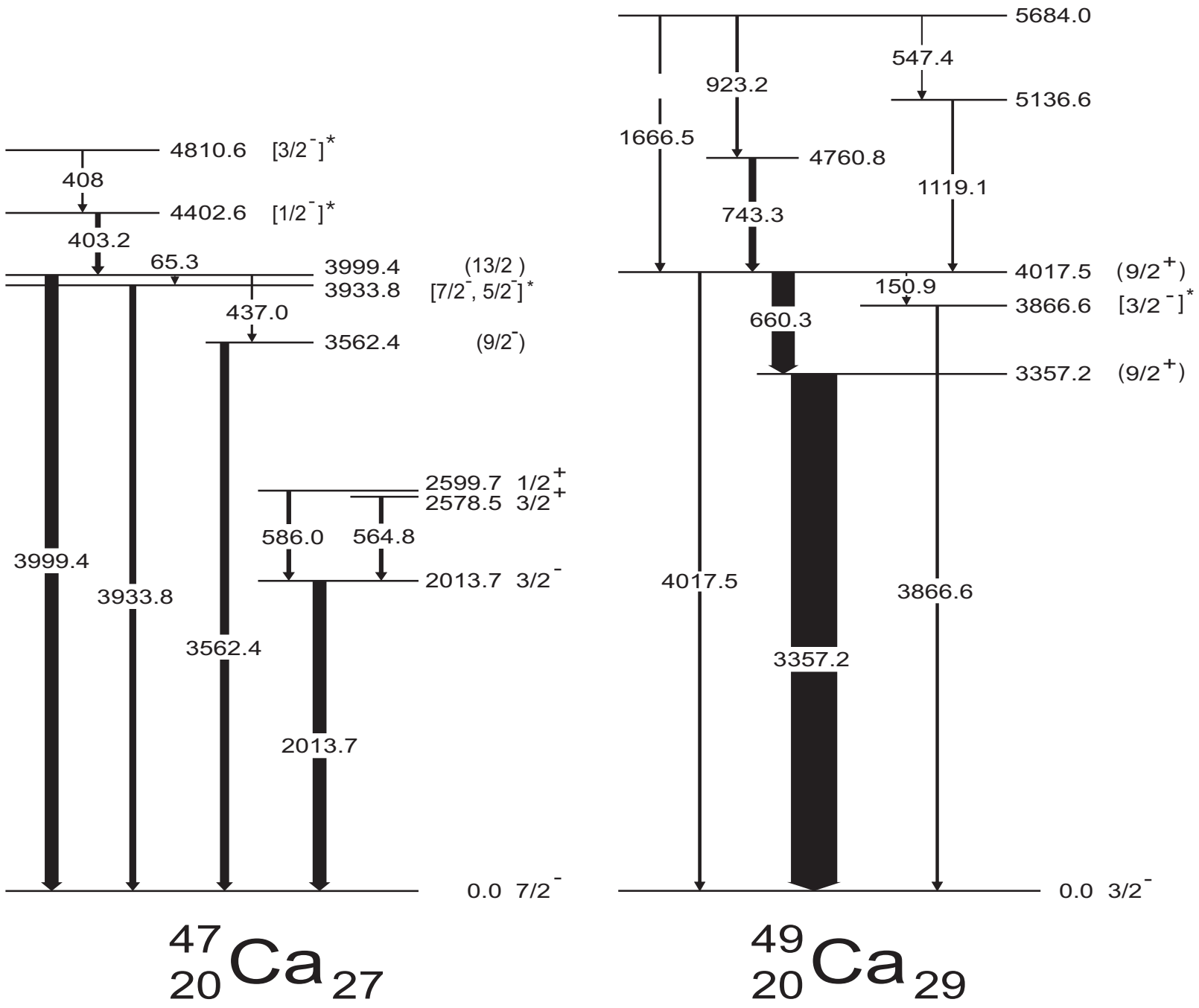

Fig. 3. Yrast levels of $^{47}$Ca and $^{49}$Ca established in the present work. Spin assignments come from the earlier study, those which are in clear conflict with present observations are placed in []* brackets.

In a similar manner we identified transitions in the one-proton $^{49}$Sc and one-proton-hole $^{47}$K nuclei for which the previous experimental knowledge was even less complete [10]. The level schemes established for both nuclei in the present work are shown in Fig. 4. Whereas in the $^{49}$Sc only two levels located at 3914 and 4046 keV energy can be attributed to previously observed states, in the $^{47}$K only the lowest $3/2^+$ state at 360 keV was known earlier.

In summary of this part we emphasize the tentative nature of presented results. We also restrained ourselves from more detailed speculations on spin-parity assignments and interpretation, which will be more justified when support from the quantitative shell model calculations is available. The independent effort to obtain such assignments from experiment is highly desirable.

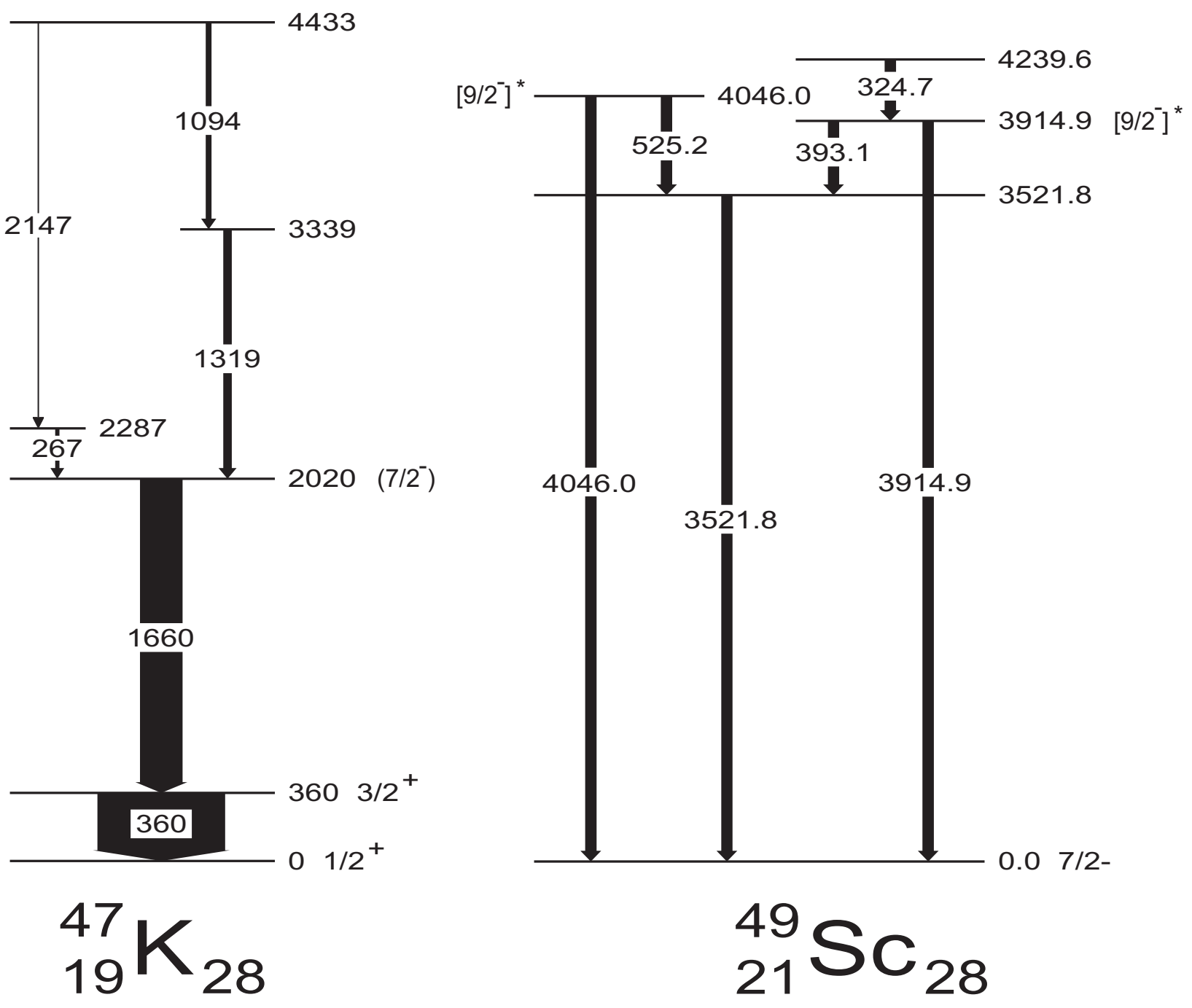

Fig. 4. Yrast levels of $^{49}$Sc and $^{47}$K established in the present work. Spin assignments as in Fig. 3.

## 3. Valence proton structures in $N$=82 isotones of the $^{132}$Sn region studied in spontaneous fission and in deep-inelastic heavy-ion reactions

Within a series of N=82 isotones the neutron-rich end close to the doubly-magic $^{132}$Sn nucleus received particular attention in recent years. The gamma coincidence measurements performed with large germanium detector arrays gave access to study yrast structures of most interesting isotones produced in the spontaneous fission of $^{248}$Cm or $^{252}$Cf nuclei. The detailed structures revealed recently for the two- and three-valence proton $^{134}$Te and $^{135}$I isotopes [8,11] provided basic knowledge of empirical two-body interactions used in the shell model calculations. In an earlier comprehensive shell model analysis, which included all available data in the series of N=82 isotones, T Wildenthal constructed the full Hamiltonian [12] which described consistently observed yrast structures. With the new experimen-tal input coming from the study of the $^{134}$Te, $^{135}$I and $^{136}$Xe isotones [13] J.Blomqvist made further refinment of this Hamiltonian. One of important results of this effort was the detailed quantitative prediction of the yrast structure of the five-valence proton $^{137}$Cs isotope, which at the time was completely unknown experimentally. The $^{137}$Cs cannot be reached in any fusion evaporation reaction and its extremely small production yield in the spontaneous fission process contributes to difficulties in spectroscopic study of this isotope. We used the data from our GAMMASPHERE coincidence experiment $^{136}$Xe +$^{232}$Th [4], where the $^{137}$Cs could be easily identified as a product of deep-inelastic one-proton transfer reaction. The detailed analysis revealed a very transparent scheme of the $^{137}$Cs yrast levels extending to the 5.5 MeV excitation energy and maximum spin value of 31/2 [14]. Except for the two states, which could be attributed to the neutron core excitations, all other states reflected one-to-one correspondence with yrast levels calculated theoretically by considering only the coupling of five valence protons. The spectacular quantitative agreement of the experimental excitation energies with the ones calculated prior to experiment, gave clear-cut confirmation of the suggested spin-parity assignments and demonstrated the predictive power of the semi-empirical shell model approach. The high intrinsic accuracy of such truncated shell model is illustrated by the root mean square deviation between the experimental and calculated energies, which was established to be less then 23 keV.

The detailed information on the neutron-rich N=82 isotones study may be found in references quoted in this section. Spontaneous fission spectroscopy provided also many other interesting results on high-spin states in simple shell model nuclei located in the region of the doubly-magic$^{132}$Sn nucleus. [eg.15,16].

## 4. New high-spin states in the $^{208}$Pb core and in the two-proton hole $^{206}$Hg nucleus

The use of deep-inelastic heavy-ion reactions in the gamma spectroscopy study gave also access to new yrast structures of many nuclei located in the region of the doubly-magic $^{208}$Pb [17,18,19]. Already in the early stage of our thick target gamma coincidence experiments several new high spin states could be observed in the $^{208}$Pb itself [7]. They were uniquely identified as simple particle-hole excitations, which extended up to the highest spin $14^-$ state at 6.744 keV, arising from the coupling of the $j_{15/2}$ neutron particle with the $i_{13/2}$ neutron hole. The continued experimental effort identified even higher spin states including the 28 ns isomeric state of yet unknown origin and nicely completed the level scheme below the 14- state [20].

In the data obtained from the new experiment $^{48}$Ca +$^{208}$Pb described already in Sec.2 rather favorable population of high-spin states in the $^{208}$Pb nucleus was observed. The sorting of triple gamma coincidence events which involved two-prompt and one-delayed (100 to 600 ns) gamma transition allowed clean selection of prompt coincidences between transitions located above the $10^+$ 500 ns isomer in $^{208}$Pb. The gamma projection spectrum of the matrix obtained by setting gates on four most intense transitions below the $10^+$isomer is shown in Fig. 5. Apart from the well known transitions arising from the $^{48}$Ca, $^{47}$Ca, $^{46}$,Ca and $^{45}$Ca reaction partner products this spectrum represents the cleanest and highest statistics data for the $^{208}$Pb high-spin states, exceeding in quality anything we obtained before. Presently the analysis is in progress and I may only briefly announce our expectations based on results obtained in this initial phase. Below the $14^-$state even more detailed structure was found, including the expected $12^-$ state of the $\nu j_{15/2} i^{-1}_{13/2}$ multiplet family. As compared to the earlier knowledge, above the $14^-$ level much more complex level scheme emerges, which is more in line with expectations based on the shell model calculations. The presence of the 28 ns isomer is confirmed and its placement as well as its decay is much better established. Moreover new high spin states are identified above the 28 ns isomer and the range of the presently observed spin-value may be estimated as extending up to I=26. The spin-parity assignments and more precise shell model calculations pose challenging questions for future interpretation of the observed new structures.

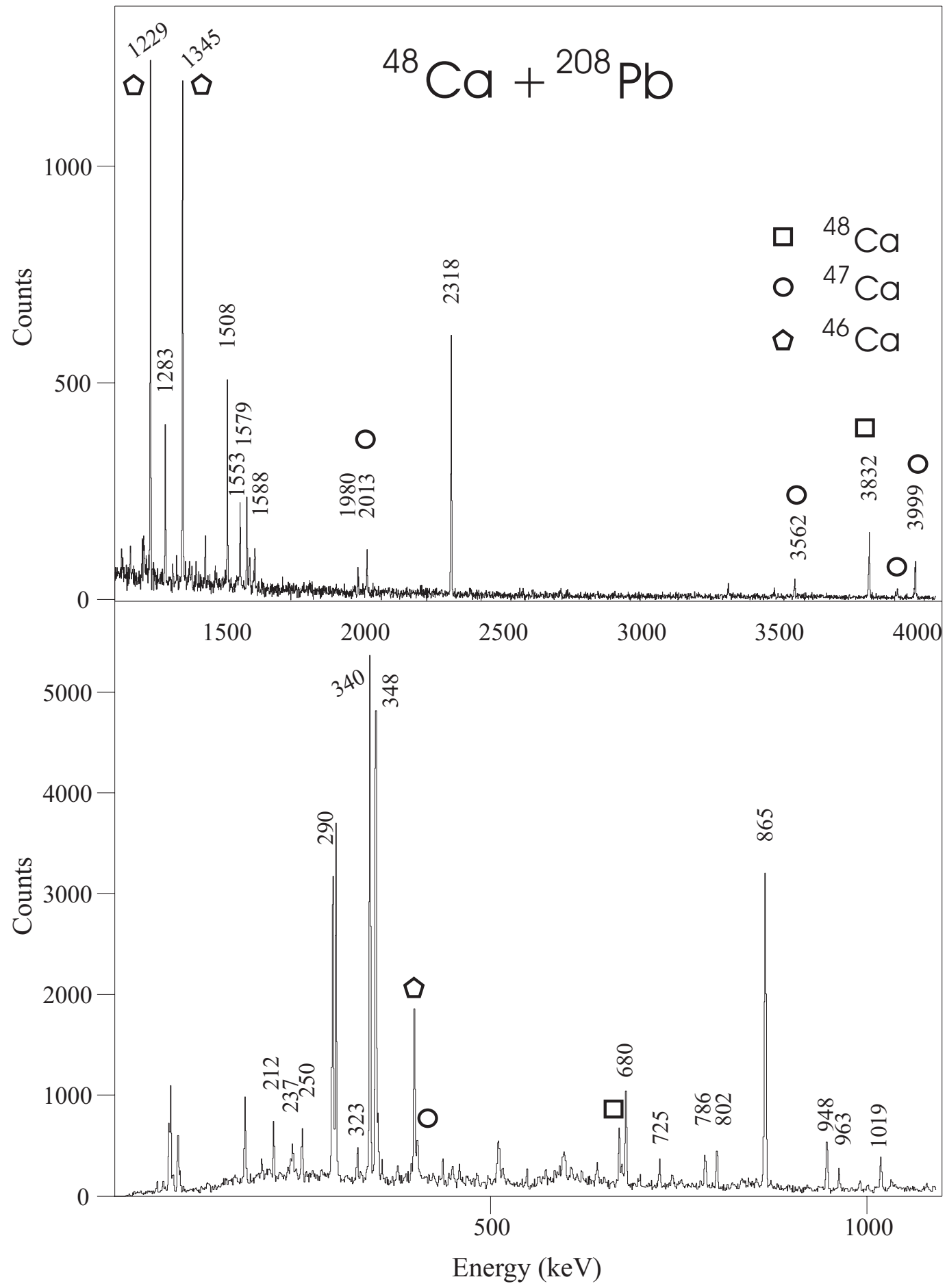

Fig. 5. Gamma projection spectrum of prompt coincidences for transitions located above the $10^+$ isomeric state in the $^{208}$Pb.

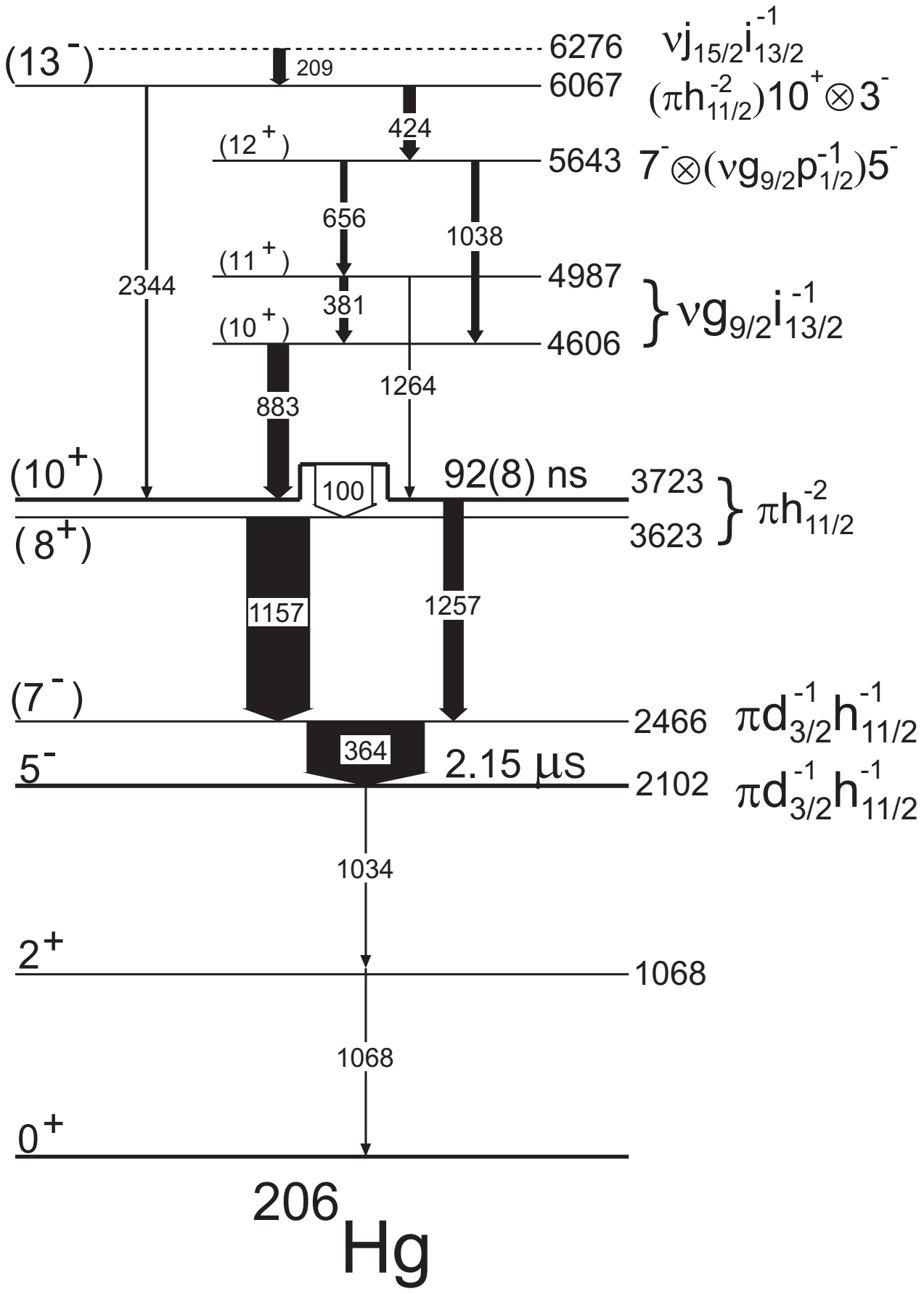

Fig. 6. Yrast level scheme of the 206Hg taken from [21]. The structural interpretation of observed levels is indicated.

The brief summary of new results obtained in our study of the $^{206}$Hg nucleus [21] is probably an appropriate and attractive topic to close my review of high-spin state investigations in the regions of doubly-magic nuclei. The long-lived 2.2 $\mu$s isomeric state $5^-$ of $\pi($ $h^{-1}_{11/2}s^{-1}_{1/2})$ structure was for long time the highest spin state known in the $^{206}$Hg isotope. In many of our experiments involving the $^{208}$Pb target we observed the population of this isomer; yet, any efforts to identify higher spin states failed. The search was mainly focussed on the identification of the expected $10^+$isomeric state of particularly pure $\pi h^{-2}_{11/2}$ structure. The $^{208}$Pb+$^{238}$U experiment performed at the ATLAS accelerator at the Argonne NL with the GAMMASPHERE array made our search successful. The 1.6 $\mu$s repetition time of the pulsed beam gave a satisfactory time space to detect delayed coincidences across the $5^-$ isomer and to identify crucial transitions of the $^{206}$Hg preceding the isomer. The complete analysis established the $^{206}$Hg level scheme which is shown in Fig. 6. The expected $10^+$isomer was unambiguously located and its 92(8) ns half-life allowed to determine the effective charge of 1.60(7)e for the $h_{11/2}$ proton hole. The subsequent analysis of delayed coincidences with transitions below this isomer identified several new states of yet higher spins. Rather straightforward interpretation of these new states is indicated in Fig. 6. The suggested 13- assignment of the 6067 keV level seems particularly interesting, since this state could be yet another example of the discussed recently [22] core octupole vibrational states, this time coupled to the two aligned $h_{11/2}$ proton holes.

## 5. Conclusion

The deep-inelastic heavy ion reactions used in thick target gamma coincidence experiments opened an access to spectroscopic study of high-spin states in the doubly-magic nuclei and their closest neighbors. We reviewed briefly main results obtained recently with this technique in regions of $^{48}$Ca, $^{132}$Sn and $^{208}$Pb doubly-closed shell nuclei, including few completely fresh and tentative results. Particularly those new results might be challenging to theory and experiment. Dedicated theoretical calculations, as well as experimental efforts aiming at spin-parity assignments should help to clarify the interpretation of observed yrast structures.

The author expresses his sincere gratitude to all collaborators who contributed to any of the presented results. The new tentative findings included in this presentation will demand further cooperative effort leading to future publications validating these results. The authors group involves:

B. Fornal, J. Wrzesiński, W.Królas, and T. Pawłat – H. Niewodniczański Institute of Nuclear Physics, Kraków, Poland,

D. Bazzacco, S. Lunardi, G. de Angelis, M. Cinausero, N. Marginean, C. Ur, G. Viesti – INFN Padova University and LNL Legnaro, Italy,

R.V.F. Janssens, M. Carpenter, I. Wiedenhover, D. Seweryniak – Argonne NL, Illinois, USA

P.J. Daly, P. Bhattacharyya, Z.W. Grabowski – Purdue University, W. Lafayette, Indiana, USA,

J. Gerl – GSI Darmstadt, Germany.

The present work was supported by the Polish State Scientific Committee under grant no. 2P03B-074-18.

## REFERENCES

[1] R.Broda et al., Phys.Rev.Lett. **68**, 1671 (1992).
[2] T.Pawat et al., Nucl.Phys. **A574**, 623 (1994).
[3] B.Fornal et al., Phys.Rev. **C55**, 762 (1997).
[4] J.F.C.Cocks et al., Phys.Rev.Lett. **78**, 2920 (1997).
[5] R.Broda et al., Phys.Rev.Lett. **74**, 868 (1995).
[6] M.Schramm et al., Z.Phys **A344**, 121 (1992).
[7] M.Schramm et al., Z.Phys. **A344**, 363 (1993).
[8] C.T.Zhang et al., Phys.Rev.Lett.**77**, 3743 (1996).
[9] Nucl.Data Sheets, **68, 1** (1993).
[10] Nucl.Data Sheets, **74, 1** (1995) and **76, 2** (1995).
[11] J.P.Omtvedt et al., Phys.Rev.Lett. **75**, 3090 (1995).
[12] B.H.Wildenthal, in *Understanding the Variety of Nuclear Excitations*, edited by A.Covello (world Scientific, Singapore, 1991).
[13] P.J.Daly et al., Phys.Rev. **C 59**, 3066 (1999).
[14] R.Broda et al., Phys.Rev. **C 59**, 3071 (1999).
[15] P.Bhattacharyya et al. Phys.Rev. **C 56**, R2363 (1997).
[16] B.Fornal et al., Phys.Rev. **C 63**, 024322 (2001).
[17] M.Rejmund et al., Z.Phys., **A 359**, 243 (1997).
[18] M.Rejmund et al., Eur. Phys. J. **A 1**, 261 (1998).
[19] B.Fornal et al., Eur. Phys. J. A 1, 355 (1998).
[20] R.Broda et al., *Proceedings of the Conf. on Nuclear Structure at the Limits, Argonne, Illinois,* July 1996, ANL/PHY –97/1, p.276 and J.Wrzesiski et al., Eur. Phys.J. A *in print.*
[21] B.Fornal et al., to be published.
[22] M.Rejmund et al., Eur. Phys. J. **A 8**, 161 (2000).